\documentclass[10pt, aps,prc,twocolumn,superscriptaddress,preprintnumbers,
amsmath, 
floatfix,
longbibliography,
nofootinbib
]{revtex4-1}
\usepackage[T1]{fontenc}
\usepackage[utf8x]{inputenc} 
\usepackage{adjustbox}          
\usepackage[caption=false]{subfig}
\usepackage{url}
\usepackage{color}
\usepackage{float}
\usepackage[pdftex,colorlinks=true, linkcolor = blue, citecolor=blue,urlcolor=blue, bookmarksnumbered=true, bookmarksopen=true]{hyperref}
\usepackage{longtable}
\usepackage{amsfonts}
\usepackage{wrapfig,bm,bbm}
\newcommand{\beq}{\begin{equation}}
\newcommand{\eeq}{\end{equation}}
\newcommand{\bea}{\begin{eqnarray}}
\newcommand{\eea}{\end{eqnarray}}

\begin{document}

\title{Fission Fragment Intrinsic Spins and Their Correlations} 

\author{Aurel Bulgac}
\affiliation{Department of Physics, University of Washington, Seattle, WA 98195--1560, USA}
\author{Ibrahim Abdurrahman} 
\affiliation{Department of Physics, University of Washington, Seattle, WA 98195--1560, USA}
\author{Shi Jin} 
\affiliation{Department of Physics, University of Washington, Seattle, WA 98195--1560, USA}
\author{Kyle Godbey}
\affiliation{Cyclotron Institute, Texas A\&M University, College Station, TX 77843, USA}
\author{Nicolas Schunck}
\affiliation{Nuclear and Chemical and Sciences Division, Lawrence Livermore 
National Laboratory, Livermore, CA 94551, USA}
\author{Ionel Stetcu}
\affiliation{Theoretical Division, Los Alamos National Laboratory, Los Alamos, NM 87545, USA}

\date{\today}

\begin{abstract}

The intrinsic spins and their correlations are the least understood
characteristics of fission fragments from both theoretical and experimental points of view. 
In many nuclear reactions the emerging fragments are typically excited and acquire an intrinsic excitation energy and 
an intrinsic spin depending on the type of the reactions and interaction mechanism. Both the 
intrinsic excitation energies and the fragments intrinsic spins and parities are controlled by the interaction 
mechanism and conservations laws, which lead to their correlations and determines the character of their
de-excitation mechanism. We outline here a framework for the theoretical extraction of the intrinsic spin  distributions
of the fragments and their correlations  within the fully microscopic real-time density functional theory formalism  
and illustrate it on the example of induced fission of   $^{236}$U and $^{240}$Pu, using two nuclear energy density functionals. 
These fission fragment intrinsic spin distributions display new qualitative features previously not discussed
in literature. Within this fully microscopic framework we extract for the first time the intrinsic spin 
distributions of fission fragments of $^{236}$U and $^{240}$Pu as well as the correlations of 
their intrinsic spins, which have been debated in literature for more 
than six decades with no definite conclusions so far.  

\end{abstract}

\preprint{NT@UW-20-10, LA-UR-20-30404}

\maketitle

In nuclear reactions a transient system is formed, which may reach statistical equilibrium as in the case of  Bohr's 
compound nucleus~\cite{Bohr:1936,*Bohr:1936a} or may only survive for a time shorter than that required to reach statistical equilibrium.
The nature of the transient system varies widely, depending on the nature and individual characteristics  of the colliding partners, 
their initial quantum numbers and collision energies, and the conservation laws that always control the evolution of the system and 
the nature of the final reaction products. As a rule the final products do not emerge with well defined quantum numbers 
such as particle number, intrinsic spins, isospins, parities,  linear momenta or intrinsic energies.
Understanding and being able to evaluate the mass and charge fragments yields, their final kinetic energies, 
their intrinsic excitation energy sharing mechanism,  the intrinsic spins and their correlations, and the decay mechanism
of the emerging primary products are of outmost interest
for understanding the reaction mechanism and for technological applications as well. In particular, the intrinsic energy
distributions and their intrinsic spin distributions will determine how the primary reaction or fission products de-excite and emit
various other particles. If well equilibrated fragments are produced then
well established statistical arguments can be used~\cite{Weisskopf:1937,Hauser:1952,Vogt:2012,Vogt:2013,Stetcu:2013,
Becker:2013,Randrup:2014,Stetcu:2014a,Talou:2018,Randrup:2019,CGMF:2020,Vogt:2020}.

Intrinsic spin distributions of primary fission fragments (FFs)
cannot be directly assessed in the laboratory, they control the neutron and $\gamma$-emission spectra, and 
consequently a significantly fraction of the energy released in fission.  The correlations between the intrinsic spins of the emerging primary FF
in particular has been a source of a debate, driven by models, remained unsettled for more than six 
decades~\cite{Strutinsky:1960,Huizenga:1960,Vandenbosch:1960,Nix:1965,
Rasmussen:1969,Wilhelmy:1972,Vandenbosch:1973,Moretto:1980,Dossing:1985,Moretto:1989,
Wagemans:1991,Bonneau:2007,Becker:2013,Vogt:2013a,Randrup:2014}.   
The scission mechanism is still not fully elucidated and both phenomenological models and incomplete microscopic models 
often based on  conflicting theoretical assumptions about the character of the large amplitude collective 
motion~\cite{Pomorski:2012,Schunck:2016,Bulgac:2020,Bender:2020}, 
lead to similar predictions for the fission yields distributions. 
The current implementation of the time-dependent density functional theory (TDDFT) extended to superfluid systems~\cite{Bulgac:2013a,Bulgac:2019}
has proven capable of providing answers to a wide number of problems in cold atom physics, quantum turbulence in 
fermionic superfluids, vortex dynamics in neutron star crust, nuclear fission and reactions. The  
DFT and the Schr{\" o}dinger descriptions are mathematically identical for one-body 
densities~\cite{Dreizler:1990lr,Gross:2006,Gross:2012}, with the proviso that in nuclear physics neither 
the nuclear energy density functional (NEDF) nor the inter-nucleon forces are known with sufficient accuracy yet. 

At scission (and immediately after) the FFs are still interacting and can still exchange 
energy, linear, and angular momentum~\cite{Bulgac:2020a}.  
These processes can lead to various relative excitation modes of the 
FFs known as axial rotation/tilting, twisting, wriggling, and bending,
the existence and importance of which is still of matter of mostly abstract debate, as a direct and unequivocal experimental 
proof of their existence and relevance is still lacking. Even if an experimental confirmation of their existence and relevance may prove hard to find,
a firm microscopic evidence of the existence of these modes, rooted in a fully quantum treatment may however be achieved.  We present here
a theoretical framework, which allows us to extract the FF intrinsic spin distributions and as well as their correlations, 
which can shed light for the first time on the existence and nature of these long speculated axial rotation/tilting, twisting, 
wriggling, and bending modes, with the latter two being doubly degenerate. 

\begin{figure}
\includegraphics[width=1\columnwidth]{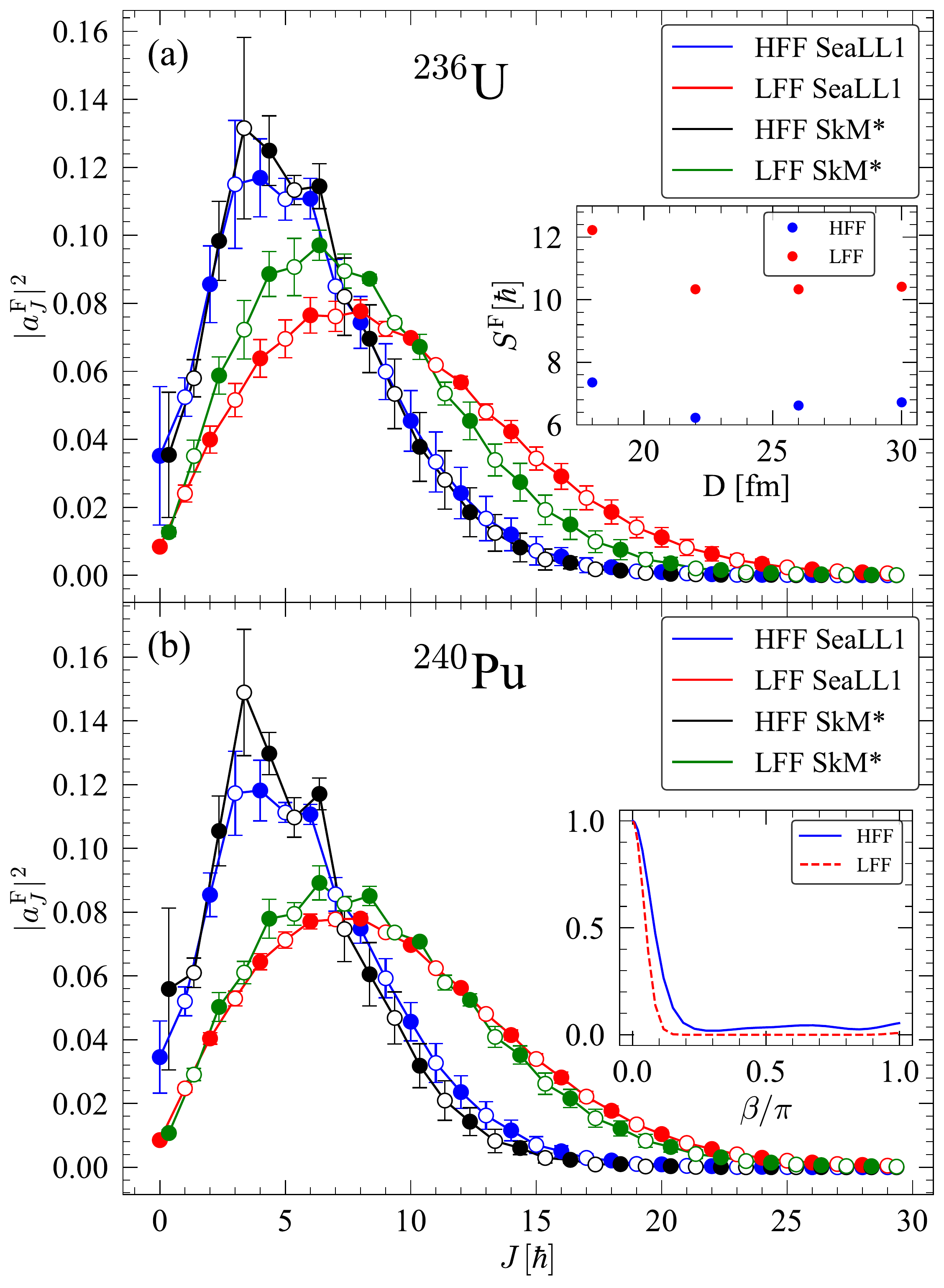}
\caption{ \label{fig:AMDs} (Color online) 
Values of $|a_J^\text{F}|^2$ averaged over initial multipole moments $Q_{20},Q_{30}$ 
from the even $J^\text{F}$-momenta are displayed with filled symbols, while the contributions arising from 
odd $J^\text{F}$-momenta are displayed with empty symbols. The ``error bars'' characterize the range of the variation 
due to the spread of initial multipole moments $Q_{20}$ and $Q_{30}$ and energies
of the fissioning nucleus.  The $|a_J^\text{F}|^2$  for the SeaLL1~\cite{Shi:2018} and SkM$^*$~\cite{Bartel:1982} 
(displaced by $\Delta J^\text{F} = 0.36$ for better visualization) NEDFs are displayed
with filled and empty symbols for the even and odd values of $J$ respectively. The average (standard deviation) for
 $^{240}$Pu are $[A^\text{L}, Z^\text{L}]$=  [103.6(0.7), 41.0(0.3)] and  $^{236}$U [102.4(2.0), 40.4(0.7)]  in case of SeaLL1 and 
[104.3(1.5), 41.4(0.5)] and  [97.9(1.2), 38.9(0.4)]  in case of  SkM$^*$ respectively.
The evaluated FF intrinsic spins $S^\text{H,L}$ at different FFs separations are shown in the inset for $^{236}U$.
Typical behavior of the overlaps $\langle \Phi | \hat{R}_x^\text{F} (\beta) | \Phi \rangle $, for one
TDDFT trajectory is shown in the inset for $^{240}$Pu. The overlaps' widths narrow with increasing $\beta$ and 
the average $S^\text{F}$ increases.}
\end{figure}

We performed TDDFT calculations of $^{236}$U and $^{240}$Pu using two different NEDFs,
SkM$^*$~\cite{Bartel:1982} and SeaLL1~\cite{Shi:2018}, in simulation boxes $30^2\times60$  with a lattice constant 
$l=1$ fm and a corresponding momentum cutoff $p_\text{cut}=\pi\hbar/l\approx 600$ MeV/c, 
using the LISE package as described in Refs.~\cite{Bulgac:2016,Bulgac:2019a,Shi:2020,IA:2020,Bulgac:2019b,Bulgac:2020}. 
The  initial nuclear wave function $\Phi$ was evolved in time from various initial deformations $Q_{20}$ and $Q_{30}$
of the mother nucleus near the outer saddle until the FFs 
were separated by more than 30 fm as in Refs~\cite{Bulgac:2019b,Bulgac:2020}. Our 
simulations have a number of significant differences from previous 
phenomenological and restricted microscopic studies available in literature. {\bf I)} There are no assumptions, 
apart from  initial axial symmetry of the fissioning nucleus, or 
restrictions imposed on the time evolution of the fissioning nucleus and of the emergent FFs. 
However, we have shown that allowing for initial states with small non-axial symmetry does not lead to 
major changes in the final properties of the FFs, see Section 3.5.3 in Ref.~\cite{Shi:2020}.
Collective rotations and shape vibrations of the mother nucleus that
contribute to quantum fluctuations  are beyond DFT~\cite{Bulgac:2019a} and are not taken into account in TDDFT.
Since the initial fissioning nucleus is deformed it also rotates, but 
with a very large rotational period $\text{T}_\text{rot}\approx 3\times10^4$ fm/c, which is much longer than the 
time the nucleus spends from saddle-to-scission $T_\text{s2s}={\cal O}(10^3)$ fm/c, 
and therefore the intrinsic nuclear shape has relatively little time 
to rotate significantly away from the fission direction.  $\text{T}_\text{rot}$ in the initial 
state can estimated from the energy of the first rotational state $2^+$ of $^{236}$U, 
$\Delta E/\Delta J\approx \hbar \omega =2\pi\hbar/\text{T}_\text{rot} \approx 40$ keV. 
Moreover, while evolving 
from the  ground state shape towards the outer fission barrier the nucleus elongates, its moment of inertia 
increases considerably and leads to an even longer rotational period. 
{\bf II)} We study the stability of 
our results with respect to varying the nuclear density functionals and the properties of the 
final FF intrinsic spin distributions appear stable.
As we stressed in our previous publications~\cite{Bulgac:2019b,Bulgac:2020}
the results of these simulations are surprisingly stable with varying the parameters of the nuclear energy 
density functionals, in good agreement with observations, without any attempts of fitting parameters.
{\bf III)} We make no assumptions about the properties of 
the emerging FFs and their ``average properties'' are noticeably different from their 
phenomenologically prescribed or equilibrium properties, 
and they are defined only after full separation.
{\bf IV)} The FF shapes  have enough time to relax, as we follow them long in time after scission and 
the FF large amplitude collective motion is strongly dissipative also. 

As soon as the FFs are well separated~\cite{Bulgac:2019b,Bulgac:2020} 
it is safe to assume that the FF intrinsic spins are not evolving anymore, 
 see Fig.~\ref{fig:AMDs}(a).
The intrinsic spin of a FF  is evaluated then  as~\cite{Sekizawa:2017a,Bulgac:2019d,Shi:2020} 
$\bm{J}^\text{F}=\int \!\!dxdy \psi^\dagger(x)\psi(y)\langle x|\bm{j}^F|y\rangle$ with
$\langle x| \bm{j}^F |y\rangle = 
\langle x|\Theta^\text{F}(\bm{r})[(\bm{r}-\bm{R}^\text{F})\times(\bm{p}-m\bm{v}^\text{F})+\bm{s}]\Theta^F(\bm{r})|y\rangle,$
and where $\int \!\!dxdy$ stands for integral over 3D spatial coordinates and sum over spin-isospin components,  
$\text{F=L, H}$  (light, heavy), $\bm{r}$ and $\bm{p}$ are the nucleon coordinate and momentum, $\bm{s}$ its spin,
$m$  the nucleon mass, $\bm{R}^\text{F}$ and $\bm{v}^\text{F}$ are 
the center of mass and the center of mass velocity of the respective FF, and $\Theta^\text{F}(\bm{r})=1$ 
only in a finite volume centered around that FF and otherwise $\Theta^\text{F}(\bm{r})\equiv 0$. 
 {In Ref.~\cite{Sekizawa:2017a} the fragment apparently 
was not brought into its own rest frame of reference.}
In Fig.~\ref{fig:AMDs} we show the extracted FF spin distributions
$|a_J^\text{F}|^2  = (2J+1)/2 \int_{0}^\pi d\beta \sin\beta  P_J(\cos\beta)
  \langle \Phi | \hat{R}_x^\text{F}(\beta)  | \Phi  \rangle$
with $ \sum_{J=0}^\infty |a_J^\text{F}|^2 =1$, $\hat{R}_x^\text{F}(\beta)=\exp(-i\hat{J}_x^\text{F}\beta/\hbar)$, 
$P_J(\cos \beta)$ the Legendre polynomials, and assuming that $z$ is the fission direction~\cite{Ring:2004,Bertsch:2019,Bulgac:2019d}. 
Like the initial state, the FFs have axial symmetry in our simulations. 
The presence of the projection on the FF spatial region and on its own reference frame is formally equivalent to introducing 
a reduced density matrix, when evaluating the entanglement entropy~\cite{Bulgac:2019d}.

\begin{table}
 \begin{tabular}{llrrrr}\hline \hline
Nucleus& NEDF&$S^\text{L}$& $S^\text{H}$ &$\beta^\text{L}_2$ &$\beta^\text{H}_2$   \\
\hline
$^{236}$U  & SeaLL1   & 10.5 (0.6)& 6.8(0.7) &0.67(0.07)&0.09(0.04)\\ 
$^{236}$U  & SkM$^*$ & 8.6(0.6)& 6.3(0.7) &0.46(0.10)&0.09(0.03)\\
$^{240}$Pu & SeaLL1   & 10.4(0.3) & 6.7(0.5) &0.62(0.04)&0.06(0.03)\\ 
$^{240}$Pu & SkM$^*$ & 9.4(0.4)& 5.8(0.5) &0.54(0.06)&0.06(0.03)\\
\hline \hline
\end{tabular}
\caption{\label{table:A} The averages (standard deviations) of $S^\text{F}$ and of
 $\beta^\text{F}_{2}$ are evaluated over the set of initial conditions, where 
 for each FF $\beta_{\lambda}^\text{F}= 4\pi \int\!d^3\!r \,n^\text{F}(\bm{r}) r^\lambda 
 Y_{\lambda0}(\hat{\bm{r}}) /\left [3A(1.2A^{1/3})^\lambda\right ]$, 
 where $n^\text{F}(\bm{r})$ is the FF intrinsic density.
 The FF $ \beta^\text{L}_3=0.00\, ...\, 0.02(0.02\, ...\, 0.07)$ and $\beta^\text{H}_3 =-0.09\, ... \, -0.04(0.01\, ...  \, 0.03)$ are noticeably smaller. }
\end{table}

There are a number of new qualitative aspects in 
our results when compared to previous either phenomenological or restricted 
microscopic studies~\cite{Strutinsky:1960,Huizenga:1960,Vandenbosch:1960,Nix:1965,
Rasmussen:1969,Wilhelmy:1972,Vandenbosch:1973,Moretto:1980,Dossing:1985,Moretto:1989,
Wagemans:1991,Bonneau:2007,Becker:2013,Vogt:2013a,Randrup:2014}. 
Notice that the spins $\bm{J}^F$ in Fig.~\ref{fig:AMDs} are not restricted to even 
values of $J$, as in Ref.~\cite{Bertsch:2019}. 
In the absence of reflection symmetry and/or in the presence of currents the overlap lacks the  
symmetry $\langle \Phi | \hat{R}_x^\text{F} (\beta) | \Phi \rangle = \langle \Phi | \hat{R}_x^\text{F} (\pi - \beta) | \Phi \rangle$, and thus
for  odd $J$-values $|a_J|^2\neq 0$. 
(Note that for a spherical nucleus $\langle \Phi | \hat{R}_x (\beta) | \Phi \rangle\equiv 1$, $|a_0|^2\equiv 1$ and $|a_{J\neq0}|^2\equiv 0$.)
This is reflected in the aspect of the overlap 
$\langle \Phi | \hat{R}_x^\text{F} (\beta) | \Phi \rangle$, which has a prominent peak at $\beta = 0$ and an almost Gaussian 
shape, see inset in Fig.~\ref{fig:AMDs}(b).
As \textcite{Scamps:2018} have noticed and was also observed by us~\cite{Bulgac:2016,Bulgac:2019b,Bulgac:2020} in 
independent calculations with different NEDFs and different implementation of TDDFT, 
FFs emerge with non-vanishing octupole deformations. 
The light FFs (LFFs) are extremely elongated when the FFs are well separated with 
$\beta_2^\text{L}\approx 5\dots 10\beta_2^\text{H}$, see Table~\ref{table:A}. 
It is not surprising that the open shell LFFs have
large deformations and thus can sustain quite large collective angular momenta, unlike the heavy FFs (HFFs).
For decades in literature it was stated that the mass and charge of the HFF
is correlated with its proximity to the magic nucleus, typically $^{132}$Sn or 
$^{208}$Pb in the case of fission of superheavy elements,
and with a strong role of the shell effects~\cite{Strutinsky:1967,BRACK:1972}. 
Since the HFFs are 
always close to the magic $^{132}$Sn nucleus their deformations are smaller than those of the LFFs, a fact reflected 
in the character of the overlaps $\langle \Phi | \hat{R}_x^F (\beta) | \Phi \rangle$ and by the 
evaluated primary FFs spins.  This is at odds with 
phenomenological inferences that the HFFs can carry a larger intrinsic spin, and doubts about the 
veracity of such an assumption were raised for quite some time~\cite{Wagemans:1991}.
At large separations the octupole moments of both FFs are relatively small, $\beta_3^\text{L}= 0.00\ldots 0.02(0.02\ldots0.07)$
and $\beta_3^\text{H}= -0.05\dots0.09(0.01\ldots0.03)$. 
The maximum and the range of the collective spin
a nucleus can sustain are larger for more deformed nuclei~\cite{Bohr:1969,Ring:2004}.

There are clear odd-even $J$-effects in the $|a_J|^2$ distributions, and the odd values of $J$ are 
slightly suppressed when compared to the neighboring even values of $J$. One should remember that we did not 
perform FF particle number projections and these odd-even effects appear for the ``average even-even'' FFs.
The distributions $|a_J^\text{H}|^2$ of the HFF 
show a prominent two peak structure. An additional feature is a rather prominent enhancement of the average 
value of $P(0)$ in case of the HFF,  larger than expected value of $|a_0|^2$, 
when compared to a statistical approach distribution~\cite{Huizenga:1960,Vandenbosch:1960,Nix:1965,
Rasmussen:1969,Wilhelmy:1972,Vandenbosch:1973,Wagemans:1991},
and also as seen from the significant ``error bar'' of $|a_0|^2$.
The gross features of the
spin distributions obtained within TDDFT, see Fig.~\ref{fig:AMDs}, can be reasonably well 
reproduced with phenomenological/statistical approach  formula $|a_J|^2\propto (2J+1)\exp\left [-{J(J+1)}/{2\sigma^2}\right ],$
where $\sigma$ is typically a fitting parameter.
For each set of initial conditions $Q_{20},Q_{30}$, as described in 
Refs.~\cite{Bulgac:2016,Bulgac:2019a,Shi:2020,IA:2020,Bulgac:2019b,Bulgac:2020},
we have extracted the values of $S^F$ for each FF from the corresponding 
$|a_J^\text{F}|^2$ distribution $S^\text{F}(S^\text{F}+1)=\sum_JJ(J+1)|a_J^\text{F}|^2,$
and then we evaluated their averages and standard deviation over the initial conditions $Q_{20},Q_{30}$, see Table~\ref{table:A}.
The SeaLL1 NEDF leads to a bit wider spin distributions than the SkM$^*$ NEDF, but otherwise
to comparable widths. The even-odd effects are more pronounced in the case of SkM$^*$ NEDF and 
particularly in the case of LFF, due likely to its reduced effective nucleon mass, 
and emerges with a noticeable octupole deformation~\cite{Bulgac:2016,Scamps:2018,Bulgac:2019b,Bulgac:2020,IA:2020}. 

\begin{table}
 \begin{tabular}{llrrr}\hline \hline
Nucleus& NEDF& $\langle \Phi|J_x^\text{L}J_x^\text{H}|\Phi\rangle $ &  
$\langle \Phi|J_y^\text{L}J_y^\text{H}|\Phi\rangle $  &  $\langle \Phi|J_z^\text{L}J_z^\text{H}|\Phi \rangle $    \\
\hline
$^{236}$U  & SeaLL1   & -1.16(0.63)& -1.16(0.63) & -2.63(0.47)\\ 
$^{236}$U  & SkM$^*$ & -0.48(0.71)& -0.48(0.71) & -1.62(0.30)\\ 
$^{240}$Pu & SeaLL1   & -0.72(0.65)& -0.72(0.65) & -4.43(0.92)\\ 
$^{240}$Pu & SkM$^*$ & -0.90(0.57)& -0.90(0.57) & -1.80(0.52)\\ 
\hline \hline
\end{tabular}
\caption{\label{table:B} The averages (standard deviations) 
of $\langle \Phi | J_\alpha^\text{L}J_\alpha^\text{H} | \Phi\rangle$, with $\alpha = x,y, z$. 
The non-diagonal elements of this tensor are negligible and all
$\langle \Phi | J^F_\alpha|\Phi\rangle = 0$. 
}
\end{table}

The correlation between the intrinsic spins of two FFs 
$\langle \Phi | J_\alpha^\text{L} J_\beta^\text{H} | \Phi\rangle = \langle \Phi | J_\beta^\text{H} J_\alpha^\text{L} | \Phi\rangle$
reveals information about the FF dynamics at and after scission.
By determining the principal axes of the tensor $\langle \Phi | J_\alpha^\text{L} J_\beta^\text{H} | \Phi\rangle$ 
and the corresponding eigenvalues one can disentangle and characterize  the 
relevance of the axial rotation/tilting, wriggling, twisting, and bending 
modes~\cite{Strutinsky:1960,Nix:1965,Moretto:1980,Dossing:1985,Moretto:1989,Vogt:2013a,Randrup:2014,Vogt:2020}.
Since $\langle \Phi|J_\alpha^\text{L}J_\alpha^\text{H}|\Phi\rangle  <0$ we confirm the presence of the  
bending and twisting modes in fission, with the bending mode being double degenerate, as expected.   
These conclusions are based for the first time on a detailed microscopic description of the fission
process in a quantum mechanical real-time many-body treatment, without any assumptions 
and no restrictions at the mean field level,  in contradistinction 
with previous phenomenological models or restricted microscopic studies. We cannot exclude however the presence
to some (small) admixture of axial rotation/tilting and wriggling, corresponding to FF rotations 
around the fission direction and perpendicular to the fission direction respectively, likely due to fluctuations and/or presence of $K\neq 0$ components.

It is instructive to qualitatively analyze these results in the semiclassical limit.
From data in Tables~\ref{table:A} and \ref{table:B} it follows that the FF intrinsic spins are on average
orthogonal to each other, as the value of cosine of their angle is small
$\cos \phi^\text{LH} = { \langle \Phi | \bm{J}^\text{L}\cdot\bm{J}^\text{H}|\Phi\rangle }/{J^\text{L}J^\text{H}}\approx 0.1.$
(For two random vectors the cosine would be $0\pm1/\sqrt{3}$.) 
As the total angular momentum is conserved  $\bm{J}_0=\bm{J}^\text{L}+\bm{J}^\text{H}+\bm{L}$ and $J^\text{L,H}_z=L_z=0$,
these angular momenta are all approximately perpendicular to the fission direction $z$. 
After introducing the total intrinsic  FF spin $\bm{J}=\bm{J}^\text{L}+\bm{J}^\text{H}$  one finds that  $J\approx 12...13$.  
At $E_n'\approx 20$ MeV, according to the analysis performed in Ref.~\cite{Vogt:2020} 
in case of $^{235}$U(n,f) the angular momentum brought in by the neutron  
can reach $\approx 5\hbar$,  and thus $J_0$ can reach values comparable,  to $J$ and $L$. 
As the ground state spins of $^{239}$Pu and $^{235}$U are $1/2^+$ and $7/2^-$, 
for slow neutrons the spins of the compound nuclei formed in $^{239}$Pu(n,f) and $^{235}$U(n,f) reactions are 
$J_0$($^{240}$Pu) = $0^+,1^+$ and  $J_0$($^{236}$U) = $3^-,4^-$, with $J_0$ noticeably smaller than $J$ and $L$.
Since the rotation of the fission direction
is controlled by the moment of inertia ${\cal I}_R= M^\text{H}M^\text{L}R^2/(M^\text{H}+M^\text{L})\rightarrow \infty$, 
where $M^\text{L,H}$ are the FF masses and $R$ their  separation, this rotation angle is expected to be relatively small.     

\begin{figure}
\includegraphics[width=1.1\columnwidth]{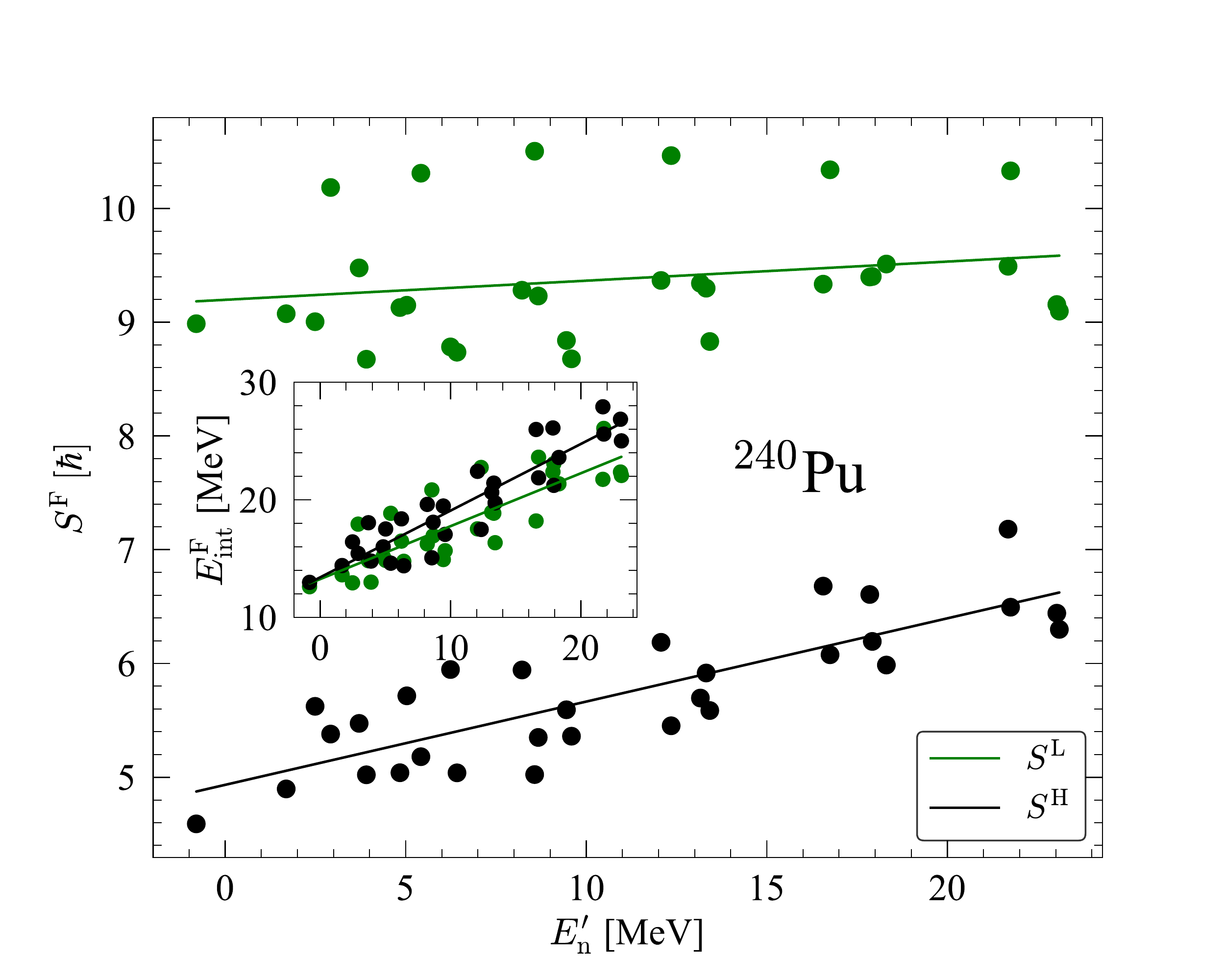}
\caption{ \label{fig:EvsJ} (Color online) 
The average intrinsic spins $S^\text{L,H}$ versus the initial FF equivalent incident neutron energy $E_n'= E^*-S_n$ 
($E^*$ and $S_n$ are the excitation energy and $S_n$ the neutron separation energy)
for the reaction 
$^{239}$Pu(n,f) with SkM$^*$ NEDF. The solid 
lines are linear fits over the data, $S^\text{L}=0.0168\,E_n'+9.197$ and $S^\text{H}=0.0732\, E_n'+4.933$ respectively,
as a function of equivalent neutron energy $E_n'$ along with their linear fits.
In the inset we display the FF excitation energies and their linear fits $E^\text{L}_\text{int}=0.4505 \, E_n'+ 13.25$ and 
$E^\text{H}_\text{int}=0.5676\,E_n'+13.40$. Using 
$E^\text{F}_\text{int}\approx A^\text{F}(T^\text{F})^2/10$~\cite{Bulgac:2019b,Bohr:1969} it follows that on average $T^\text{L}>T^\text{H}$. }
\end{figure}

In the case of $^{240}$Pu we have performed additional simulations with the NEDF SkM$^*$ by varying the equivalent incident neutron 
energy in reaction $^{239}$Pu(n,f), thus simulating a compound nucleus $^{240}$Pu 
with various excitation energies $E^*$, see Fig.~\ref{fig:EvsJ}. 
With increasing $E_n$ the intrinsic spin of the HFF shows a significant increase,
which correlates with the steeper increase of the HFF excitation energy $E^\text{H}_\text{int}$
when compared to the behavior of the LFF excitation energy $E^\text{L}_\text{int}$. Nevertheless, in 
this entire energy interval the HFF ``temperature'' remains lower than the LFF ``temperature'' on average.

A recent constrained Hartree-Fock-Bogoliubov evaluation of the FF intrinsic spins~\cite{Schunck:2020x}, using
pre-scicisson configurations with the same $A^\text{L}$ and $Z^\text{L}$ as the final FF values obtained in 
TDDFT calculations from different initial conditions~\cite{Bulgac:2019d}, arrived at similar results to these reported here, 
provided the neck thickness at rupture is chosen small enough. As the FF deformations and excitation energies change significantly
after scission~\cite{Bulgac:2019b,Bulgac:2020} and the FF moments of inertia which are $\propto\beta^2_2$~\cite{Ring:2004},
the intrinsic spin distributions change with FF separation, see Fig.~\ref{fig:AMDs}a.
We have compared the default CGMF results~\cite{Becker:2013,Stetcu:2013,Stetcu:2014a,Talou:2018,CGMF:2020} 
for the $\gamma$-spectra with those obtained by using instead the microscopic intrinsic 
spin distributions.
While fewer average number of gammas were produced when the microscopic parameterization 
was employed, we have not observed large changes for the prompt fission $\gamma$-spectrum.
One should keep in mind that the CGMF model 
is based on a large number of phenomenological parameters. In
the CGMF approach one assumes that $J^\text{H}>J^\text{L}$, opposite to our conclusions. 
This assumption is hard to reconcile with the fact that the HFF  has  a relatively modest deformation. 
In another study~\cite{Vogt:2020}, within the phenomenological model FREYA  one
finds that FF intrinsic antiparallel intrinsic spins show a slight preference. Using our language,
the expected average of $\langle \Phi|J_\alpha^\text{L}J_\alpha^\text{H}|\Phi\rangle $ for $\alpha = x,y$  is negative, in qualitative agreement with our results. 
In FREYA the intrinsic spin fluctuations are controlled by the temperature of the  fissioning nucleus at scission.  According to our 
earlier results~\cite{Bulgac:2019b,Bulgac:2020} the FF deformations at scission and their relaxed values are very different, with larger 
deformations at scission and with the HFF cooler than the LFF, while FREYA assumes identical temperatures. 
While within FREYA framework the ``thermal'' intrinsic spin 
fluctuations dominate over their averages and these authors find that average 
difference $|S^\text{H}-S^\text{L}| \approx 1\ldots 2$ is smaller than our values. 
Since in FREYA the FF moments of inertia are ${\cal I}^\text{H}>{\cal I}^\text{L}$ it immediately follows that 
$S^\text{H}>S^\text{L}$, opposite to  our results.  At the same time  FREYA finds $S^\text{H}$ values
and an increase in  $S^\text{L,H}$ with the excitation energy of the compound 
fissioning nucleus qualitatively similar to our findings, see Fig.~\ref{fig:EvsJ}.
In FREYA the axial rotation/tilting and twisting modes are suppressed~\cite{Dossing:1985}, while we find that
magnitude of $\langle \Phi|J_z^\text{L}J_z^\text{H}|\Phi\rangle $ is larger than $\langle \Phi|J_\alpha^\text{L}J_\alpha^\text{H}|\Phi\rangle$ for $\alpha = x,y$.

We have demonstrated that TDDFT allows one to extract detailed microscopic information 
about the FF intrinsic spins, their dependence on
excitation energy of the compound nucleus, and the FF intrinsic spins correlations, which 
are almost impossible to infer unambiguously from phenomenological analyses.

\begin{acknowledgements}
AB thanks G.~F. Bertsch for discussions. We also want to express our 
gratitude to K.~J. Roche for enthusiastically sharing with us 
his insights into effectively using supercomputers. 

AB devised the theoretical framework.
IA, SJ,  and IS performed TDDFT calculations and IA, KG, 
and IS implemented and performed the extraction of the spin distributions. 
NS performed the calculations of the initial configurations for the TDDFT simulations. 
All authors participated in the discussion of the results and the writing of the manuscript.  

AB was supported by U.S. Department of Energy,
Office of Science, Grant No. DE-FG02-97ER41014.  and in part by NNSA
cooperative Agreement DE-NA0003841.
The work of NS was supported by the Scientific Discovery through
Advanced Computing (SciDAC) program funded by the U.S.~Department of
Energy, Office of Science, Advanced Scientific Computing Research and
Nuclear Physics, and it was partly performed under the auspices of the
US Department of Energy by the Lawrence Livermore National Laboratory
under Contract DE-AC52-07NA27344.  
The work of IS was supported by the US Department of Energy through the 
Los Alamos National Laboratory. Los Alamos National Laboratory is operated 
by Triad National Security, LLC, for the National Nuclear Security Administration 
of U.S. Department of Energy (Contract No. 89233218CNA000001). 
IS gratefully acknowledges partial support by the Laboratory Directed Research 
and Development program of Los Alamos National Laboratory under project 
number 20200384ER and partial support and computational resources 
provided by the Advanced Simulation and Computing (ASC) Program.
This material (work of AB, IA, and KG) is partially based upon work supported by the Department of 
Energy, National Nuclear Security Administration, under Award Number DE-NA0003841.

Some of the calculations reported
here have been performed with computing support from the Lawrence
Livermore National Laboratory (LLNL) Institutional Computing Grand
Challenge program.
This research used resources of the Oak Ridge
Leadership Computing Facility, which is a U.S. DOE Office of Science
User Facility supported under Contract No. DE-AC05-00OR22725 and of
the National Energy Research Scientific computing Center, which is
supported by the Office of Science of the U.S. Department of Energy
under Contract No. DE-AC02-05CH11231.
We acknowledge PRACE for
awarding us access to resource Piz Daint based at the Swiss National
Supercomputing Centre (CSCS), decision No. 2018194657.
This work is supported by ``High Performance Computing 
Infrastructure'' in Japan, Project ID: hp180048. A series of simulations 
were carried out on the Tsubame 3.0 supercomputer at Tokyo Institute of Technology.
This research used resources provided by the Los Alamos National Laboratory 
Institutional Computing Program.
\end{acknowledgements}


\providecommand{\selectlanguage}[1]{}
\renewcommand{\selectlanguage}[1]{}

\bibliography{latest_fission}

\begin{thebibliography}{53}%
\makeatletter
\providecommand \@ifxundefined [1]{%
 \@ifx{#1\undefined}
}%
\providecommand \@ifnum [1]{%
 \ifnum #1\expandafter \@firstoftwo
 \else \expandafter \@secondoftwo
 \fi
}%
\providecommand \@ifx [1]{%
 \ifx #1\expandafter \@firstoftwo
 \else \expandafter \@secondoftwo
 \fi
}%
\providecommand \natexlab [1]{#1}%
\providecommand \enquote  [1]{``#1''}%
\providecommand \bibnamefont  [1]{#1}%
\providecommand \bibfnamefont [1]{#1}%
\providecommand \citenamefont [1]{#1}%
\providecommand \href@noop [0]{\@secondoftwo}%
\providecommand \href [0]{\begingroup \@sanitize@url \@href}%
\providecommand \@href[1]{\@@startlink{#1}\@@href}%
\providecommand \@@href[1]{\endgroup#1\@@endlink}%
\providecommand \@sanitize@url [0]{\catcode `\\12\catcode `\$12\catcode
  `\&12\catcode `\#12\catcode `\^12\catcode `\_12\catcode `\%12\relax}%
\providecommand \@@startlink[1]{}%
\providecommand \@@endlink[0]{}%
\providecommand \url  [0]{\begingroup\@sanitize@url \@url }%
\providecommand \@url [1]{\endgroup\@href {#1}{\urlprefix }}%
\providecommand \urlprefix  [0]{URL }%
\providecommand \Eprint [0]{\href }%
\providecommand \doibase [0]{http://dx.doi.org/}%
\providecommand \selectlanguage [0]{\@gobble}%
\providecommand \bibinfo  [0]{\@secondoftwo}%
\providecommand \bibfield  [0]{\@secondoftwo}%
\providecommand \translation [1]{[#1]}%
\providecommand \BibitemOpen [0]{}%
\providecommand \bibitemStop [0]{}%
\providecommand \bibitemNoStop [0]{.\EOS\space}%
\providecommand \EOS [0]{\spacefactor3000\relax}%
\providecommand \BibitemShut  [1]{\csname bibitem#1\endcsname}%
\let\auto@bib@innerbib\@empty
\bibitem [{\citenamefont {Bohr}(1936)}]{Bohr:1936}%
  \BibitemOpen
  \bibfield  {author} {\bibinfo {author} {\bibfnamefont {N.}~\bibnamefont
  {Bohr}},\ }\bibfield  {title} {\enquote {\bibinfo {title} {{Neutron Capture
  and Nuclear Constitution}},}\ }\href {\doibase 10.1038/137344a0} {\bibfield
  {journal} {\bibinfo  {journal} {Nature}\ }\textbf {\bibinfo {volume} {137}},\
  \bibinfo {pages} {344} (\bibinfo {year} {1936})}\BibitemShut {NoStop}%
\bibitem [{\citenamefont {{Nature Editors}}(1936)}]{Bohr:1936a}%
  \BibitemOpen
  \bibfield  {author} {\bibinfo {author} {\bibnamefont {{Nature Editors}}},\
  }\bibfield  {title} {\enquote {\bibinfo {title} {{Neutron Capture and Nuclear
  Constitution}},}\ }\href {\doibase 10.1038/137351a0} {\bibfield  {journal}
  {\bibinfo  {journal} {Nature}\ }\textbf {\bibinfo {volume} {137}},\ \bibinfo
  {pages} {351} (\bibinfo {year} {1936})}\BibitemShut {NoStop}%
\bibitem [{\citenamefont {Weisskopf}(1937)}]{Weisskopf:1937}%
  \BibitemOpen
  \bibfield  {author} {\bibinfo {author} {\bibfnamefont {V.}~\bibnamefont
  {Weisskopf}},\ }\bibfield  {title} {\enquote {\bibinfo {title} {{Statistics
  and Nuclear Reactions}},}\ }\href {\doibase 10.1103/PhysRev.52.295}
  {\bibfield  {journal} {\bibinfo  {journal} {Phys. Rev.}\ }\textbf {\bibinfo
  {volume} {52}},\ \bibinfo {pages} {295} (\bibinfo {year} {1937})}\BibitemShut
  {NoStop}%
\bibitem [{\citenamefont {Hauser}\ and\ \citenamefont
  {Feshbach}(1952)}]{Hauser:1952}%
  \BibitemOpen
  \bibfield  {author} {\bibinfo {author} {\bibfnamefont {W.}~\bibnamefont
  {Hauser}}\ and\ \bibinfo {author} {\bibfnamefont {H.}~\bibnamefont
  {Feshbach}},\ }\bibfield  {title} {\enquote {\bibinfo {title} {The inelastic
  scattering of neutrons},}\ }\href {\doibase 10.1103/PhysRev.87.366}
  {\bibfield  {journal} {\bibinfo  {journal} {Phys. Rev.}\ }\textbf {\bibinfo
  {volume} {87}},\ \bibinfo {pages} {366} (\bibinfo {year} {1952})}\BibitemShut
  {NoStop}%
\bibitem [{\citenamefont {Vogt}\ \emph {et~al.}(2012)\citenamefont {Vogt},
  \citenamefont {Randrup}, \citenamefont {Brown}, \citenamefont {Descalle},\
  and\ \citenamefont {Ormand}}]{Vogt:2012}%
  \BibitemOpen
  \bibfield  {author} {\bibinfo {author} {\bibfnamefont {R.}~\bibnamefont
  {Vogt}}, \bibinfo {author} {\bibfnamefont {J.}~\bibnamefont {Randrup}},
  \bibinfo {author} {\bibfnamefont {D.~A.}\ \bibnamefont {Brown}}, \bibinfo
  {author} {\bibfnamefont {M.~A.}\ \bibnamefont {Descalle}}, \ and\ \bibinfo
  {author} {\bibfnamefont {W.~E.}\ \bibnamefont {Ormand}},\ }\bibfield  {title}
  {\enquote {\bibinfo {title} {Event-by-event evaluation of the prompt fission
  neutron spectrum from ${}^{239}\text{Pu}(n,f)$},}\ }\href {\doibase
  10.1103/PhysRevC.85.024608} {\bibfield  {journal} {\bibinfo  {journal} {Phys.
  Rev. C}\ }\textbf {\bibinfo {volume} {85}},\ \bibinfo {pages} {024608}
  (\bibinfo {year} {2012})}\BibitemShut {NoStop}%
\bibitem [{\citenamefont {Vogt}\ and\ \citenamefont
  {Randrup}(2013{\natexlab{a}})}]{Vogt:2013}%
  \BibitemOpen
  \bibfield  {author} {\bibinfo {author} {\bibfnamefont {R.}~\bibnamefont
  {Vogt}}\ and\ \bibinfo {author} {\bibfnamefont {J.}~\bibnamefont {Randrup}},\
  }\bibfield  {title} {\enquote {\bibinfo {title} {{Event-by-event Modeling of
  Prompt Neutrons and Photons from Neutron-induced and Spontaneous Fission with
  FREYA}},}\ }\href {\doibase 10.1016/j.phpro.2013.06.013} {\bibfield
  {journal} {\bibinfo  {journal} {Physics Procedia}\ }\textbf {\bibinfo
  {volume} {47}},\ \bibinfo {pages} {82} (\bibinfo {year}
  {2013}{\natexlab{a}})}\BibitemShut {NoStop}%
\bibitem [{\citenamefont {Stetcu}\ \emph {et~al.}(2013)\citenamefont {Stetcu},
  \citenamefont {Talou}, \citenamefont {Kawano},\ and\ \citenamefont
  {Jandel}}]{Stetcu:2013}%
  \BibitemOpen
  \bibfield  {author} {\bibinfo {author} {\bibfnamefont {I.}~\bibnamefont
  {Stetcu}}, \bibinfo {author} {\bibfnamefont {P.}~\bibnamefont {Talou}},
  \bibinfo {author} {\bibfnamefont {T.}~\bibnamefont {Kawano}}, \ and\ \bibinfo
  {author} {\bibfnamefont {M.}~\bibnamefont {Jandel}},\ }\bibfield  {title}
  {\enquote {\bibinfo {title} {Isomer production ratios and the angular
  momentum distribution of fission fragments},}\ }\href {\doibase
  10.1103/PhysRevC.88.044603} {\bibfield  {journal} {\bibinfo  {journal} {Phys.
  Rev. C}\ }\textbf {\bibinfo {volume} {88}},\ \bibinfo {pages} {044603}
  (\bibinfo {year} {2013})}\BibitemShut {NoStop}%
\bibitem [{\citenamefont {Becker}\ \emph {et~al.}(2013)\citenamefont {Becker},
  \citenamefont {Talou}, \citenamefont {Kawano}, \citenamefont {Danon},\ and\
  \citenamefont {Stetcu}}]{Becker:2013}%
  \BibitemOpen
  \bibfield  {author} {\bibinfo {author} {\bibfnamefont {B.}~\bibnamefont
  {Becker}}, \bibinfo {author} {\bibfnamefont {P.}~\bibnamefont {Talou}},
  \bibinfo {author} {\bibfnamefont {T.}~\bibnamefont {Kawano}}, \bibinfo
  {author} {\bibfnamefont {Y.}~\bibnamefont {Danon}}, \ and\ \bibinfo {author}
  {\bibfnamefont {I.}~\bibnamefont {Stetcu}},\ }\bibfield  {title} {\enquote
  {\bibinfo {title} {{Monte Carlo Hauser-Feshbach predictions of prompt fission
  $\ensuremath{\gamma}$ rays: Application to $n_\mathrm{th}+^{235}$U,
  $n_\mathrm{th}+^{239}$Pu, and $^{252}$Cf (sf)}},}\ }\href {\doibase
  10.1103/PhysRevC.87.014617} {\bibfield  {journal} {\bibinfo  {journal} {Phys.
  Rev. C}\ }\textbf {\bibinfo {volume} {87}},\ \bibinfo {pages} {014617}
  (\bibinfo {year} {2013})}\BibitemShut {NoStop}%
\bibitem [{\citenamefont {Randrup}\ and\ \citenamefont
  {Vogt}(2014)}]{Randrup:2014}%
  \BibitemOpen
  \bibfield  {author} {\bibinfo {author} {\bibfnamefont {J.}~\bibnamefont
  {Randrup}}\ and\ \bibinfo {author} {\bibfnamefont {R.}~\bibnamefont {Vogt}},\
  }\bibfield  {title} {\enquote {\bibinfo {title} {Refined treatment of angular
  momentum in the event-by-event fission model freya},}\ }\href {\doibase
  10.1103/PhysRevC.89.044601} {\bibfield  {journal} {\bibinfo  {journal} {Phys.
  Rev. C}\ }\textbf {\bibinfo {volume} {89}},\ \bibinfo {pages} {044601}
  (\bibinfo {year} {2014})}\BibitemShut {NoStop}%
\bibitem [{\citenamefont {Stetcu}\ \emph {et~al.}(2014)\citenamefont {Stetcu},
  \citenamefont {Talou}, \citenamefont {Kawano},\ and\ \citenamefont
  {Jandel}}]{Stetcu:2014a}%
  \BibitemOpen
  \bibfield  {author} {\bibinfo {author} {\bibfnamefont {I.}~\bibnamefont
  {Stetcu}}, \bibinfo {author} {\bibfnamefont {P.}~\bibnamefont {Talou}},
  \bibinfo {author} {\bibfnamefont {T.}~\bibnamefont {Kawano}}, \ and\ \bibinfo
  {author} {\bibfnamefont {M.}~\bibnamefont {Jandel}},\ }\bibfield  {title}
  {\enquote {\bibinfo {title} {Properties of prompt-fission
  $\ensuremath{\gamma}$ rays},}\ }\href {\doibase 10.1103/PhysRevC.90.024617}
  {\bibfield  {journal} {\bibinfo  {journal} {Phys. Rev. C}\ }\textbf {\bibinfo
  {volume} {90}},\ \bibinfo {pages} {024617} (\bibinfo {year}
  {2014})}\BibitemShut {NoStop}%
\bibitem [{\citenamefont {Talou}\ \emph {et~al.}(2018)\citenamefont {Talou},
  \citenamefont {Vogt}, \citenamefont {Randrup}, \citenamefont {Rising},
  \citenamefont {Pozzi}, \citenamefont {Verbeke}, \citenamefont {Andrews},
  \citenamefont {Clarke}, \citenamefont {Jaffke}, \citenamefont {Jandel},
  \citenamefont {Kawano}, \citenamefont {Marcath}, \citenamefont
  {Meierbachtol}, \citenamefont {Nakae}, \citenamefont {Rusev}, \citenamefont
  {Sood}, \citenamefont {Stetcu},\ and\ \citenamefont {Walker}}]{Talou:2018}%
  \BibitemOpen
  \bibfield  {author} {\bibinfo {author} {\bibfnamefont {P.}~\bibnamefont
  {Talou}}, \bibinfo {author} {\bibfnamefont {R.}~\bibnamefont {Vogt}},
  \bibinfo {author} {\bibfnamefont {J.}~\bibnamefont {Randrup}}, \bibinfo
  {author} {\bibfnamefont {M.~E.}\ \bibnamefont {Rising}}, \bibinfo {author}
  {\bibfnamefont {S.~A.}\ \bibnamefont {Pozzi}}, \bibinfo {author}
  {\bibfnamefont {J.}~\bibnamefont {Verbeke}}, \bibinfo {author} {\bibfnamefont
  {M.~T.}\ \bibnamefont {Andrews}}, \bibinfo {author} {\bibfnamefont {S.~D.}\
  \bibnamefont {Clarke}}, \bibinfo {author} {\bibfnamefont {P.}~\bibnamefont
  {Jaffke}}, \bibinfo {author} {\bibfnamefont {M.}~\bibnamefont {Jandel}},
  \bibinfo {author} {\bibfnamefont {T.}~\bibnamefont {Kawano}}, \bibinfo
  {author} {\bibfnamefont {M.~J.}\ \bibnamefont {Marcath}}, \bibinfo {author}
  {\bibfnamefont {K.}~\bibnamefont {Meierbachtol}}, \bibinfo {author}
  {\bibfnamefont {L.}~\bibnamefont {Nakae}}, \bibinfo {author} {\bibfnamefont
  {G.}~\bibnamefont {Rusev}}, \bibinfo {author} {\bibfnamefont
  {A.}~\bibnamefont {Sood}}, \bibinfo {author} {\bibfnamefont {I.}~\bibnamefont
  {Stetcu}}, \ and\ \bibinfo {author} {\bibfnamefont {C.}~\bibnamefont
  {Walker}},\ }\bibfield  {title} {\enquote {\bibinfo {title} {Correlated
  prompt fission data in transport simulations},}\ }\href {\doibase
  10.1140/epja/i2018-12455-0} {\bibfield  {journal} {\bibinfo  {journal} {Eur.
  Phys. Jour. A}\ }\textbf {\bibinfo {volume} {54}},\ \bibinfo {pages} {9}
  (\bibinfo {year} {2018})}\BibitemShut {NoStop}%
\bibitem [{\citenamefont {Randrup}\ \emph {et~al.}(2019)\citenamefont
  {Randrup}, \citenamefont {Talou},\ and\ \citenamefont {Vogt}}]{Randrup:2019}%
  \BibitemOpen
  \bibfield  {author} {\bibinfo {author} {\bibfnamefont {J.}~\bibnamefont
  {Randrup}}, \bibinfo {author} {\bibfnamefont {P.}~\bibnamefont {Talou}}, \
  and\ \bibinfo {author} {\bibfnamefont {R.}~\bibnamefont {Vogt}},\ }\bibfield
  {title} {\enquote {\bibinfo {title} {Sensitivity of neutron observables to
  the model input in simulations of ${}^{252}\mathrm{Cf}(\mathrm{sf})$},}\
  }\href {\doibase 10.1103/PhysRevC.99.054619} {\bibfield  {journal} {\bibinfo
  {journal} {Phys. Rev. C}\ }\textbf {\bibinfo {volume} {99}},\ \bibinfo
  {pages} {054619} (\bibinfo {year} {2019})}\BibitemShut {NoStop}%
\bibitem [{\citenamefont {Talou}\ \emph {et~al.}()\citenamefont {Talou},
  \citenamefont {Stetcu}, \citenamefont {Jafke}, \citenamefont {Rising},
  \citenamefont {Lovell},\ and\ \citenamefont {Kawano}}]{CGMF:2020}%
  \BibitemOpen
  \bibfield  {author} {\bibinfo {author} {\bibfnamefont {P.}~\bibnamefont
  {Talou}}, \bibinfo {author} {\bibfnamefont {I.}~\bibnamefont {Stetcu}},
  \bibinfo {author} {\bibfnamefont {P.}~\bibnamefont {Jafke}}, \bibinfo
  {author} {\bibfnamefont {M.~E.}\ \bibnamefont {Rising}}, \bibinfo {author}
  {\bibfnamefont {A.~E}\ \bibnamefont {Lovell}}, \ and\ \bibinfo {author}
  {\bibfnamefont {T.}~\bibnamefont {Kawano}},\ }\href@noop {} {\enquote
  {\bibinfo {title} {{Cascading Gamma-ray Multiplicity and Fission, open source
  code: github.com/lanl/cgmf}},}\ }\BibitemShut {NoStop}%
\bibitem [{\citenamefont {Vogt}\ and\ \citenamefont
  {Randrup}(2021)}]{Vogt:2020}%
  \BibitemOpen
  \bibfield  {author} {\bibinfo {author} {\bibfnamefont {R.}~\bibnamefont
  {Vogt}}\ and\ \bibinfo {author} {\bibfnamefont {J.}~\bibnamefont {Randrup}},\
  }\bibfield  {title} {\enquote {\bibinfo {title} {Angular momentum effects in
  fission},}\ }\href {\doibase 10.1103/PhysRevC.103.014610} {\bibfield
  {journal} {\bibinfo  {journal} {Phys. Rev. C}\ }\textbf {\bibinfo {volume}
  {103}},\ \bibinfo {pages} {014610} (\bibinfo {year} {2021})}\BibitemShut
  {NoStop}%
\bibitem [{\citenamefont {Strutinsky}(1960)}]{Strutinsky:1960}%
  \BibitemOpen
  \bibfield  {author} {\bibinfo {author} {\bibfnamefont {V.~M.}\ \bibnamefont
  {Strutinsky}},\ }\bibfield  {title} {\enquote {\bibinfo {title} {{Angular
  Anisotropy of Gamma Quanta that Accompany Fission}},}\ }\href
  {http://jetp.ac.ru/cgi-bin/e/index/e/10/3/p613?a=list} {\bibfield  {journal}
  {\bibinfo  {journal} {Sov. Phys. JETP}\ }\textbf {\bibinfo {volume} {10}},\
  \bibinfo {pages} {613} (\bibinfo {year} {1960})}\BibitemShut {NoStop}%
\bibitem [{\citenamefont {Huizenga}\ and\ \citenamefont
  {Vandenbosch}(1960)}]{Huizenga:1960}%
  \BibitemOpen
  \bibfield  {author} {\bibinfo {author} {\bibfnamefont {J.~R.}\ \bibnamefont
  {Huizenga}}\ and\ \bibinfo {author} {\bibfnamefont {R.}~\bibnamefont
  {Vandenbosch}},\ }\bibfield  {title} {\enquote {\bibinfo {title}
  {Interpretation of isomeric cross-section ratios for ($n,
  \ensuremath{\gamma}$) and ($\ensuremath{\gamma}, n$) reactions},}\ }\href
  {\doibase 10.1103/PhysRev.120.1305} {\bibfield  {journal} {\bibinfo
  {journal} {Phys. Rev.}\ }\textbf {\bibinfo {volume} {120}},\ \bibinfo {pages}
  {1305--1312} (\bibinfo {year} {1960})}\BibitemShut {NoStop}%
\bibitem [{\citenamefont {Vandenbosch}\ and\ \citenamefont
  {Huizenga}(1960)}]{Vandenbosch:1960}%
  \BibitemOpen
  \bibfield  {author} {\bibinfo {author} {\bibfnamefont {R.}~\bibnamefont
  {Vandenbosch}}\ and\ \bibinfo {author} {\bibfnamefont {J.~R.}\ \bibnamefont
  {Huizenga}},\ }\bibfield  {title} {\enquote {\bibinfo {title} {Isomeric
  cross-section ratios for reactions producing the isomeric pair
  ${\mathrm{hg}}^{197,197m}$},}\ }\href {\doibase 10.1103/PhysRev.120.1313}
  {\bibfield  {journal} {\bibinfo  {journal} {Phys. Rev.}\ }\textbf {\bibinfo
  {volume} {120}},\ \bibinfo {pages} {1313--1318} (\bibinfo {year}
  {1960})}\BibitemShut {NoStop}%
\bibitem [{\citenamefont {Nix}\ and\ \citenamefont
  {Swiatecki}(1965)}]{Nix:1965}%
  \BibitemOpen
  \bibfield  {author} {\bibinfo {author} {\bibfnamefont {J.~R.}\ \bibnamefont
  {Nix}}\ and\ \bibinfo {author} {\bibfnamefont {W.~J.}\ \bibnamefont
  {Swiatecki}},\ }\bibfield  {title} {\enquote {\bibinfo {title} {Studies in
  the liquid-drop theory of nuclear fission},}\ }\href {\doibase
  https://doi.org/10.1016/0029-5582(65)90038-6} {\bibfield  {journal} {\bibinfo
   {journal} {Nucl. Phys.}\ }\textbf {\bibinfo {volume} {71}},\ \bibinfo
  {pages} {1} (\bibinfo {year} {1965})}\BibitemShut {NoStop}%
\bibitem [{\citenamefont {Rasmussen}\ \emph {et~al.}(1969)\citenamefont
  {Rasmussen}, \citenamefont {N{\"o}renberg},\ and\ \citenamefont
  {Mang}}]{Rasmussen:1969}%
  \BibitemOpen
  \bibfield  {author} {\bibinfo {author} {\bibfnamefont {J.O.}\ \bibnamefont
  {Rasmussen}}, \bibinfo {author} {\bibfnamefont {W.}~\bibnamefont
  {N{\"o}renberg}}, \ and\ \bibinfo {author} {\bibfnamefont {H.J.}\
  \bibnamefont {Mang}},\ }\bibfield  {title} {\enquote {\bibinfo {title} {A
  model for calculating the angular momentum distribution of fission
  fragments},}\ }\href {\doibase https://doi.org/10.1016/0375-9474(69)90066-9}
  {\bibfield  {journal} {\bibinfo  {journal} {Nucl. Phys. A}\ }\textbf
  {\bibinfo {volume} {136}},\ \bibinfo {pages} {465} (\bibinfo {year}
  {1969})}\BibitemShut {NoStop}%
\bibitem [{\citenamefont {Wilhelmy}\ \emph {et~al.}(1972)\citenamefont
  {Wilhelmy}, \citenamefont {Cheifetz}, \citenamefont {Jared}, \citenamefont
  {Thompson}, \citenamefont {Bowman},\ and\ \citenamefont
  {Rasmussen}}]{Wilhelmy:1972}%
  \BibitemOpen
  \bibfield  {author} {\bibinfo {author} {\bibfnamefont {J.~B.}\ \bibnamefont
  {Wilhelmy}}, \bibinfo {author} {\bibfnamefont {E.}~\bibnamefont {Cheifetz}},
  \bibinfo {author} {\bibfnamefont {R.~C.}\ \bibnamefont {Jared}}, \bibinfo
  {author} {\bibfnamefont {S.~G.}\ \bibnamefont {Thompson}}, \bibinfo {author}
  {\bibfnamefont {H.~R.}\ \bibnamefont {Bowman}}, \ and\ \bibinfo {author}
  {\bibfnamefont {J.~O.}\ \bibnamefont {Rasmussen}},\ }\bibfield  {title}
  {\enquote {\bibinfo {title} {Angular momentum of primary products formed in
  the spontaneous fission of $^{252}\mathrm{Cf}$},}\ }\href {\doibase
  10.1103/PhysRevC.5.2041} {\bibfield  {journal} {\bibinfo  {journal} {Phys.
  Rev. C}\ }\textbf {\bibinfo {volume} {5}},\ \bibinfo {pages} {2041--2060}
  (\bibinfo {year} {1972})}\BibitemShut {NoStop}%
\bibitem [{\citenamefont {Vandenbosch}\ and\ \citenamefont
  {Huizenga}(1973)}]{Vandenbosch:1973}%
  \BibitemOpen
  \bibfield  {author} {\bibinfo {author} {\bibfnamefont {R.}~\bibnamefont
  {Vandenbosch}}\ and\ \bibinfo {author} {\bibfnamefont {J.R.}\ \bibnamefont
  {Huizenga}},\ }\href@noop {} {\emph {\bibinfo {title} {{Nuclear Fission}}}}\
  (\bibinfo  {publisher} {{Academic Press}},\ \bibinfo {address} {New York},\
  \bibinfo {year} {1973})\BibitemShut {NoStop}%
\bibitem [{\citenamefont {Moretto}\ and\ \citenamefont
  {Schmitt}(1980)}]{Moretto:1980}%
  \BibitemOpen
  \bibfield  {author} {\bibinfo {author} {\bibfnamefont {L.~G.}\ \bibnamefont
  {Moretto}}\ and\ \bibinfo {author} {\bibfnamefont {R.~P.}\ \bibnamefont
  {Schmitt}},\ }\bibfield  {title} {\enquote {\bibinfo {title} {Equilibrium
  statistical treatment of angular momenta associated with collective modes in
  fission and heavy-ion reactions},}\ }\href {\doibase 10.1103/PhysRevC.21.204}
  {\bibfield  {journal} {\bibinfo  {journal} {Phys. Rev. C}\ }\textbf {\bibinfo
  {volume} {21}},\ \bibinfo {pages} {204} (\bibinfo {year} {1980})}\BibitemShut
  {NoStop}%
\bibitem [{\citenamefont {D{\o}ssing}\ and\ \citenamefont
  {Randrup}(1985)}]{Dossing:1985}%
  \BibitemOpen
  \bibfield  {author} {\bibinfo {author} {\bibfnamefont {T.}~\bibnamefont
  {D{\o}ssing}}\ and\ \bibinfo {author} {\bibfnamefont {J.}~\bibnamefont
  {Randrup}},\ }\bibfield  {title} {\enquote {\bibinfo {title} {{Dynamical
  evolution of angular momentum in damped nuclear reactions: (I). Accumulation
  of angular momentum by nucleon transfer}},}\ }\href {\doibase
  https://doi.org/10.1016/0375-9474(85)90178-2} {\bibfield  {journal} {\bibinfo
   {journal} {Nucl. Phys. A}\ }\textbf {\bibinfo {volume} {433}},\ \bibinfo
  {pages} {215} (\bibinfo {year} {1985})}\BibitemShut {NoStop}%
\bibitem [{\citenamefont {Moretto}\ \emph {et~al.}(1989)\citenamefont
  {Moretto}, \citenamefont {Peaslee},\ and\ \citenamefont
  {Wozniak}}]{Moretto:1989}%
  \BibitemOpen
  \bibfield  {author} {\bibinfo {author} {\bibfnamefont {L.~G.}\ \bibnamefont
  {Moretto}}, \bibinfo {author} {\bibfnamefont {G.~F.}\ \bibnamefont
  {Peaslee}}, \ and\ \bibinfo {author} {\bibfnamefont {G.~J.}\ \bibnamefont
  {Wozniak}},\ }\bibfield  {title} {\enquote {\bibinfo {title}
  {{Angular-Momentum-Bearing Modes in Fission}},}\ }\href {\doibase
  10.1016/0375-9474(89)90682-9} {\bibfield  {journal} {\bibinfo  {journal}
  {Nucl. Phys. A}\ }\textbf {\bibinfo {volume} {502}},\ \bibinfo {pages} {453c}
  (\bibinfo {year} {1989})}\BibitemShut {NoStop}%
\bibitem [{\citenamefont {Wagemans}(1991)}]{Wagemans:1991}%
  \BibitemOpen
  \bibinfo {editor} {\bibfnamefont {C.}~\bibnamefont {Wagemans}},\ ed.,\
  \href@noop {} {\emph {\bibinfo {title} {{The Nuclear Fission Process}}}}\
  (\bibinfo  {publisher} {CRS Press, Boca Raton},\ \bibinfo {year}
  {1991})\BibitemShut {NoStop}%
\bibitem [{\citenamefont {Bonneau}\ \emph {et~al.}(2007)\citenamefont
  {Bonneau}, \citenamefont {Quentin},\ and\ \citenamefont
  {Mikhailov}}]{Bonneau:2007}%
  \BibitemOpen
  \bibfield  {author} {\bibinfo {author} {\bibfnamefont {L.}~\bibnamefont
  {Bonneau}}, \bibinfo {author} {\bibfnamefont {P.}~\bibnamefont {Quentin}}, \
  and\ \bibinfo {author} {\bibfnamefont {I.~N.}\ \bibnamefont {Mikhailov}},\
  }\bibfield  {title} {\enquote {\bibinfo {title} {Scission configurations and
  their implication in fission-fragment angular momenta},}\ }\href {\doibase
  10.1103/PhysRevC.75.064313} {\bibfield  {journal} {\bibinfo  {journal} {Phys.
  Rev. C}\ }\textbf {\bibinfo {volume} {75}},\ \bibinfo {pages} {064313}
  (\bibinfo {year} {2007})}\BibitemShut {NoStop}%
\bibitem [{\citenamefont {Vogt}\ and\ \citenamefont
  {Randrup}(2013{\natexlab{b}})}]{Vogt:2013a}%
  \BibitemOpen
  \bibfield  {author} {\bibinfo {author} {\bibfnamefont {R.}~\bibnamefont
  {Vogt}}\ and\ \bibinfo {author} {\bibfnamefont {J.}~\bibnamefont {Randrup}},\
  }\bibfield  {title} {\enquote {\bibinfo {title} {Event-by-event study of
  photon observables in spontaneous and thermal fission},}\ }\href {\doibase
  10.1103/PhysRevC.87.044602} {\bibfield  {journal} {\bibinfo  {journal} {Phys.
  Rev. C}\ }\textbf {\bibinfo {volume} {87}},\ \bibinfo {pages} {044602}
  (\bibinfo {year} {2013}{\natexlab{b}})}\BibitemShut {NoStop}%
\bibitem [{\citenamefont {Krappe}\ and\ \citenamefont
  {Pomorski}(2012)}]{Pomorski:2012}%
  \BibitemOpen
  \bibfield  {author} {\bibinfo {author} {\bibfnamefont {J.~K.}\ \bibnamefont
  {Krappe}}\ and\ \bibinfo {author} {\bibfnamefont {K.}~\bibnamefont
  {Pomorski}},\ }\href {\doibase 10.1007/978-3-642-23515-3} {\emph {\bibinfo
  {title} {{Theory of Nuclear Fission}}}}\ (\bibinfo  {publisher} {Springer
  Heidelberg},\ \bibinfo {year} {2012})\BibitemShut {NoStop}%
\bibitem [{\citenamefont {Schunck}\ and\ \citenamefont
  {Robledo}(2016)}]{Schunck:2016}%
  \BibitemOpen
  \bibfield  {author} {\bibinfo {author} {\bibfnamefont {N.}~\bibnamefont
  {Schunck}}\ and\ \bibinfo {author} {\bibfnamefont {L.~M.}\ \bibnamefont
  {Robledo}},\ }\bibfield  {title} {\enquote {\bibinfo {title} {{Microscopic}
  theory of nuclear fission: a review},}\ }\href {\doibase
  10.1088/0034-4885/79/11/116301} {\bibfield  {journal} {\bibinfo  {journal}
  {{Rep. Prog. Phys.}}\ }\textbf {\bibinfo {volume} {79}},\ \bibinfo {pages}
  {116301} (\bibinfo {year} {2016})}\BibitemShut {NoStop}%
\bibitem [{\citenamefont {Bulgac}\ \emph {et~al.}(2020)\citenamefont {Bulgac},
  \citenamefont {Jin},\ and\ \citenamefont {Stetcu}}]{Bulgac:2020}%
  \BibitemOpen
  \bibfield  {author} {\bibinfo {author} {\bibfnamefont {A.}~\bibnamefont
  {Bulgac}}, \bibinfo {author} {\bibfnamefont {S.}~\bibnamefont {Jin}}, \ and\
  \bibinfo {author} {\bibfnamefont {I.}~\bibnamefont {Stetcu}},\ }\bibfield
  {title} {\enquote {\bibinfo {title} {{Nuclear Fission Dynamics: Past,
  Present, Needs, and Future}},}\ }\href {\doibase 10.3389/fphy.2020.00063}
  {\bibfield  {journal} {\bibinfo  {journal} {{Frontiers in Physics}}\ }\textbf
  {\bibinfo {volume} {8}},\ \bibinfo {pages} {63} (\bibinfo {year}
  {2020})}\BibitemShut {NoStop}%
\bibitem [{\citenamefont {Bender}\ \emph {et~al.}(2020)\citenamefont {Bender},
  \citenamefont {Bernard}, \citenamefont {Bertsch}, \citenamefont {Chiba},
  \citenamefont {Dobaczewski}, \citenamefont {Dubray}, \citenamefont
  {Giuliani}, \citenamefont {Hagino}, \citenamefont {Lacroix}, \citenamefont
  {Li}, \citenamefont {Magierski}, \citenamefont {Maruhn}, \citenamefont
  {Nazarewicz}, \citenamefont {Pei}, \citenamefont {P{\'{e}}ru}, \citenamefont
  {Pillet}, \citenamefont {Randrup}, \citenamefont {Regnier}, \citenamefont
  {Reinhard}, \citenamefont {Robledo}, \citenamefont {Ryssens}, \citenamefont
  {Sadhukhan}, \citenamefont {Scamps}, \citenamefont {Schunck}, \citenamefont
  {Simenel}, \citenamefont {Skalski}, \citenamefont {Stetcu}, \citenamefont
  {Stevenson}, \citenamefont {Umar}, \citenamefont {Verriere}, \citenamefont
  {Vretenar}, \citenamefont {Warda},\ and\ \citenamefont
  {{\AA}berg}}]{Bender:2020}%
  \BibitemOpen
  \bibfield  {author} {\bibinfo {author} {\bibfnamefont {M.}~\bibnamefont
  {Bender}}, \bibinfo {author} {\bibfnamefont {R.}~\bibnamefont {Bernard}},
  \bibinfo {author} {\bibfnamefont {G.}~\bibnamefont {Bertsch}}, \bibinfo
  {author} {\bibfnamefont {S.}~\bibnamefont {Chiba}}, \bibinfo {author}
  {\bibfnamefont {J.}~\bibnamefont {Dobaczewski}}, \bibinfo {author}
  {\bibfnamefont {N.}~\bibnamefont {Dubray}}, \bibinfo {author} {\bibfnamefont
  {S.~A.}\ \bibnamefont {Giuliani}}, \bibinfo {author} {\bibfnamefont
  {K.}~\bibnamefont {Hagino}}, \bibinfo {author} {\bibfnamefont
  {D.}~\bibnamefont {Lacroix}}, \bibinfo {author} {\bibfnamefont
  {Z.}~\bibnamefont {Li}}, \bibinfo {author} {\bibfnamefont {P.}~\bibnamefont
  {Magierski}}, \bibinfo {author} {\bibfnamefont {J.}~\bibnamefont {Maruhn}},
  \bibinfo {author} {\bibfnamefont {W.}~\bibnamefont {Nazarewicz}}, \bibinfo
  {author} {\bibfnamefont {J.}~\bibnamefont {Pei}}, \bibinfo {author}
  {\bibfnamefont {S.}~\bibnamefont {P{\'{e}}ru}}, \bibinfo {author}
  {\bibfnamefont {N.}~\bibnamefont {Pillet}}, \bibinfo {author} {\bibfnamefont
  {J.}~\bibnamefont {Randrup}}, \bibinfo {author} {\bibfnamefont
  {D.}~\bibnamefont {Regnier}}, \bibinfo {author} {\bibfnamefont {P.-G.}\
  \bibnamefont {Reinhard}}, \bibinfo {author} {\bibfnamefont {L.~M.}\
  \bibnamefont {Robledo}}, \bibinfo {author} {\bibfnamefont {W.}~\bibnamefont
  {Ryssens}}, \bibinfo {author} {\bibfnamefont {J.}~\bibnamefont {Sadhukhan}},
  \bibinfo {author} {\bibfnamefont {G.}~\bibnamefont {Scamps}}, \bibinfo
  {author} {\bibfnamefont {N.}~\bibnamefont {Schunck}}, \bibinfo {author}
  {\bibfnamefont {C.}~\bibnamefont {Simenel}}, \bibinfo {author} {\bibfnamefont
  {J.}~\bibnamefont {Skalski}}, \bibinfo {author} {\bibfnamefont
  {I.}~\bibnamefont {Stetcu}}, \bibinfo {author} {\bibfnamefont
  {P.}~\bibnamefont {Stevenson}}, \bibinfo {author} {\bibfnamefont
  {S.}~\bibnamefont {Umar}}, \bibinfo {author} {\bibfnamefont {M.}~\bibnamefont
  {Verriere}}, \bibinfo {author} {\bibfnamefont {D.}~\bibnamefont {Vretenar}},
  \bibinfo {author} {\bibfnamefont {M.}~\bibnamefont {Warda}}, \ and\ \bibinfo
  {author} {\bibfnamefont {S.}~\bibnamefont {{\AA}berg}},\ }\bibfield  {title}
  {\enquote {\bibinfo {title} {Future of nuclear fission theory},}\ }\href
  {\doibase 10.1088/1361-6471/abab4f} {\bibfield  {journal} {\bibinfo
  {journal} {Journal of Physics G: Nuclear and Particle Physics}\ }\textbf
  {\bibinfo {volume} {47}},\ \bibinfo {pages} {113002} (\bibinfo {year}
  {2020})}\BibitemShut {NoStop}%
\bibitem [{\citenamefont {Bulgac}(2013)}]{Bulgac:2013a}%
  \BibitemOpen
  \bibfield  {author} {\bibinfo {author} {\bibfnamefont {A.}~\bibnamefont
  {Bulgac}},\ }\bibfield  {title} {\enquote {\bibinfo {title} {{Time-Dependent
  Density Functional Theory and the Real-Time Dynamics of Fermi
  Superfluids}},}\ }\href {\doibase 10.1146/annurev-nucl-102212-170631}
  {\bibfield  {journal} {\bibinfo  {journal} {Ann. Rev. Nucl. and Part. Sci.}\
  }\textbf {\bibinfo {volume} {63}},\ \bibinfo {pages} {97} (\bibinfo {year}
  {2013})}\BibitemShut {NoStop}%
\bibitem [{\citenamefont {Bulgac}(2019{\natexlab{a}})}]{Bulgac:2019}%
  \BibitemOpen
  \bibfield  {author} {\bibinfo {author} {\bibfnamefont {A.}~\bibnamefont
  {Bulgac}},\ }\bibfield  {title} {\enquote {\bibinfo {title} {{Time-Dependent
  Density Functional Theory for Fermionic Superfluids: from Cold Atomic gases,
  to Nuclei and Neutron Star Crust}},}\ }\href {\doibase
  10.1002/pssb.201800592} {\bibfield  {journal} {\bibinfo  {journal} {Physica
  Status Solidi B}\ }\textbf {\bibinfo {volume} {256}},\ \bibinfo {pages}
  {1800592} (\bibinfo {year} {2019}{\natexlab{a}})}\BibitemShut {NoStop}%
\bibitem [{\citenamefont {Dreizler}\ and\ \citenamefont
  {{Gross}}(1990)}]{Dreizler:1990lr}%
  \BibitemOpen
  \bibfield  {author} {\bibinfo {author} {\bibfnamefont {R.~M.}\ \bibnamefont
  {Dreizler}}\ and\ \bibinfo {author} {\bibfnamefont {E.~K.~U.}\ \bibnamefont
  {{Gross}}},\ }\href {\doibase 10.1007/978-3-642-86105-5} {\emph {\bibinfo
  {title} {{Density Functional Theory: An Approach to the Quantum Many--Body
  Problem}}}}\ (\bibinfo  {publisher} {Springer-Verlag},\ \bibinfo {address}
  {Berlin},\ \bibinfo {year} {1990})\BibitemShut {NoStop}%
\bibitem [{\citenamefont {Marques}\ \emph {et~al.}(2006)\citenamefont
  {Marques}, \citenamefont {Ullrich}, \citenamefont {Nogueira}, \citenamefont
  {Rubio}, \citenamefont {Burke},\ and\ \citenamefont {{Gross}}}]{Gross:2006}%
  \BibitemOpen
  \bibinfo {editor} {\bibfnamefont {M.~A.~L.}\ \bibnamefont {Marques}},
  \bibinfo {editor} {\bibfnamefont {C.~A.}\ \bibnamefont {Ullrich}}, \bibinfo
  {editor} {\bibfnamefont {F.}~\bibnamefont {Nogueira}}, \bibinfo {editor}
  {\bibfnamefont {A.}~\bibnamefont {Rubio}}, \bibinfo {editor} {\bibfnamefont
  {K.}~\bibnamefont {Burke}}, \ and\ \bibinfo {editor} {\bibfnamefont
  {E.~K.~U.}\ \bibnamefont {{Gross}}},\ eds.,\ \href {\doibase
  10.1007/b11767107} {\emph {\bibinfo {title} {Time-Dependent Density
  Functional Theory}}},\ \bibinfo {series} {Lecture Notes in Physics}, Vol.\
  \bibinfo {volume} {706}\ (\bibinfo  {publisher} {Springer-Verlag},\ \bibinfo
  {address} {Berlin},\ \bibinfo {year} {2006})\BibitemShut {NoStop}%
\bibitem [{\citenamefont {Marques}\ \emph {et~al.}(2012)\citenamefont
  {Marques}, \citenamefont {Maitra}, \citenamefont {Nogueira}, \citenamefont
  {{Gross}},\ and\ \citenamefont {Rubio}}]{Gross:2012}%
  \BibitemOpen
  \bibinfo {editor} {\bibfnamefont {M.~A.~L.}\ \bibnamefont {Marques}},
  \bibinfo {editor} {\bibfnamefont {N.~T.}\ \bibnamefont {Maitra}}, \bibinfo
  {editor} {\bibfnamefont {F.~M.~S.}\ \bibnamefont {Nogueira}}, \bibinfo
  {editor} {\bibfnamefont {E.~K.~U.}\ \bibnamefont {{Gross}}}, \ and\ \bibinfo
  {editor} {\bibfnamefont {A.}~\bibnamefont {Rubio}},\ eds.,\ \href {\doibase
  10.1007/978-3-642-23518-4} {\emph {\bibinfo {title} {Fundamentals of
  Time-Dependent Density Functional Theory}}},\ \bibinfo {series} {Lecture
  Notes in Physics}, Vol.\ \bibinfo {volume} {837}\ (\bibinfo  {publisher}
  {Springer},\ \bibinfo {address} {Heidelberg},\ \bibinfo {year}
  {2012})\BibitemShut {NoStop}%
\bibitem [{\citenamefont {Bulgac}(2020)}]{Bulgac:2020a}%
  \BibitemOpen
  \bibfield  {author} {\bibinfo {author} {\bibfnamefont {A.}~\bibnamefont
  {Bulgac}},\ }\bibfield  {title} {\enquote {\bibinfo {title} {Fission-fragment
  excitation energy sharing beyond scission},}\ }\href {\doibase
  10.1103/PhysRevC.102.044609} {\bibfield  {journal} {\bibinfo  {journal}
  {Phys. Rev. C}\ }\textbf {\bibinfo {volume} {102}},\ \bibinfo {pages}
  {044609} (\bibinfo {year} {2020})}\BibitemShut {NoStop}%
\bibitem [{\citenamefont {Bulgac}\ \emph {et~al.}(2018)\citenamefont {Bulgac},
  \citenamefont {Forbes}, \citenamefont {Jin}, \citenamefont {Perez},\ and\
  \citenamefont {Schunck}}]{Shi:2018}%
  \BibitemOpen
  \bibfield  {author} {\bibinfo {author} {\bibfnamefont {A.}~\bibnamefont
  {Bulgac}}, \bibinfo {author} {\bibfnamefont {M.~M.}\ \bibnamefont {Forbes}},
  \bibinfo {author} {\bibfnamefont {S.}~\bibnamefont {Jin}}, \bibinfo {author}
  {\bibfnamefont {R.~N.}\ \bibnamefont {Perez}}, \ and\ \bibinfo {author}
  {\bibfnamefont {N.}~\bibnamefont {Schunck}},\ }\bibfield  {title} {\enquote
  {\bibinfo {title} {Minimal nuclear energy density functional},}\ }\href
  {\doibase 10.1103/PhysRevC.97.044313} {\bibfield  {journal} {\bibinfo
  {journal} {Phys. Rev. C}\ }\textbf {\bibinfo {volume} {97}},\ \bibinfo
  {pages} {044313} (\bibinfo {year} {2018})}\BibitemShut {NoStop}%
\bibitem [{\citenamefont {Bartel}\ \emph {et~al.}(1982)\citenamefont {Bartel},
  \citenamefont {Quentin}, \citenamefont {Brack}, \citenamefont {Guet},\ and\
  \citenamefont {H{\aa}kansson}}]{Bartel:1982}%
  \BibitemOpen
  \bibfield  {author} {\bibinfo {author} {\bibfnamefont {J.}~\bibnamefont
  {Bartel}}, \bibinfo {author} {\bibfnamefont {P.}~\bibnamefont {Quentin}},
  \bibinfo {author} {\bibfnamefont {M.}~\bibnamefont {Brack}}, \bibinfo
  {author} {\bibfnamefont {C.}~\bibnamefont {Guet}}, \ and\ \bibinfo {author}
  {\bibfnamefont {H.-B.}\ \bibnamefont {H{\aa}kansson}},\ }\bibfield  {title}
  {\enquote {\bibinfo {title} {{Towards a better parametrisation of Skyrme-like
  effective forces: A critical study of the SkM force}},}\ }\href {\doibase
  https://doi.org/10.1016/0375-9474(82)90403-1} {\bibfield  {journal} {\bibinfo
   {journal} {Nucl. Phys. A}\ }\textbf {\bibinfo {volume} {386}},\ \bibinfo
  {pages} {79} (\bibinfo {year} {1982})}\BibitemShut {NoStop}%
\bibitem [{\citenamefont {Bulgac}\ \emph {et~al.}(2016)\citenamefont {Bulgac},
  \citenamefont {Magierski}, \citenamefont {Roche},\ and\ \citenamefont
  {Stetcu}}]{Bulgac:2016}%
  \BibitemOpen
  \bibfield  {author} {\bibinfo {author} {\bibfnamefont {A.}~\bibnamefont
  {Bulgac}}, \bibinfo {author} {\bibfnamefont {P.}~\bibnamefont {Magierski}},
  \bibinfo {author} {\bibfnamefont {K.~J.}\ \bibnamefont {Roche}}, \ and\
  \bibinfo {author} {\bibfnamefont {I.}~\bibnamefont {Stetcu}},\ }\bibfield
  {title} {\enquote {\bibinfo {title} {{Induced Fission of $^{240}\mathrm{Pu}$
  within a Real-Time Microscopic Framework}},}\ }\href {\doibase
  10.1103/PhysRevLett.116.122504} {\bibfield  {journal} {\bibinfo  {journal}
  {Phys. Rev. Lett.}\ }\textbf {\bibinfo {volume} {116}},\ \bibinfo {pages}
  {122504} (\bibinfo {year} {2016})}\BibitemShut {NoStop}%
\bibitem [{\citenamefont {Bulgac}\ \emph
  {et~al.}(2019{\natexlab{a}})\citenamefont {Bulgac}, \citenamefont {Jin},\
  and\ \citenamefont {Stetcu}}]{Bulgac:2019a}%
  \BibitemOpen
  \bibfield  {author} {\bibinfo {author} {\bibfnamefont {A.}~\bibnamefont
  {Bulgac}}, \bibinfo {author} {\bibfnamefont {S.}~\bibnamefont {Jin}}, \ and\
  \bibinfo {author} {\bibfnamefont {I.}~\bibnamefont {Stetcu}},\ }\bibfield
  {title} {\enquote {\bibinfo {title} {Unitary evolution with fluctuations and
  dissipation},}\ }\href {\doibase 10.1103/PhysRevC.100.014615} {\bibfield
  {journal} {\bibinfo  {journal} {Phys. Rev. C}\ }\textbf {\bibinfo {volume}
  {100}},\ \bibinfo {pages} {014615} (\bibinfo {year}
  {2019}{\natexlab{a}})}\BibitemShut {NoStop}%
\bibitem [{\citenamefont {Jin}\ \emph {et~al.}()\citenamefont {Jin},
  \citenamefont {Roche}, \citenamefont {Stetcu}, \citenamefont {Abdurrahman},\
  and\ \citenamefont {Bulgac}}]{Shi:2020}%
  \BibitemOpen
  \bibfield  {author} {\bibinfo {author} {\bibfnamefont {S.}~\bibnamefont
  {Jin}}, \bibinfo {author} {\bibfnamefont {K.~J.}\ \bibnamefont {Roche}},
  \bibinfo {author} {\bibfnamefont {I.}~\bibnamefont {Stetcu}}, \bibinfo
  {author} {\bibfnamefont {I}~\bibnamefont {Abdurrahman}}, \ and\ \bibinfo
  {author} {\bibfnamefont {A.}~\bibnamefont {Bulgac}},\ }\href@noop {}
  {\enquote {\bibinfo {title} {{The LISE package: solvers for static and
  time-dependent superfluid local density approximation equations in three
  dimensions}},}\ }\Eprint {http://arxiv.org/abs/2009.00745} {arXiv:2009.00745}
  \BibitemShut {NoStop}%
\bibitem [{\citenamefont {Abdurahman}\ \emph {et~al.}(2020)\citenamefont
  {Abdurahman}, \citenamefont {Bulgac}, \citenamefont {Schunck},\ and\
  \citenamefont {Stetcu}}]{IA:2020}%
  \BibitemOpen
  \bibfield  {author} {\bibinfo {author} {\bibfnamefont {I.}~\bibnamefont
  {Abdurahman}}, \bibinfo {author} {\bibfnamefont {A.}~\bibnamefont {Bulgac}},
  \bibinfo {author} {\bibfnamefont {N.}~\bibnamefont {Schunck}}, \ and\
  \bibinfo {author} {\bibfnamefont {I.}~\bibnamefont {Stetcu}},\ }\href@noop {}
  {\enquote {\bibinfo {title} {{Fission fragment properties (unpublished)}},}\
  } (\bibinfo {year} {2020})\BibitemShut {NoStop}%
\bibitem [{\citenamefont {Bulgac}\ \emph
  {et~al.}(2019{\natexlab{b}})\citenamefont {Bulgac}, \citenamefont {Jin},
  \citenamefont {Roche}, \citenamefont {Schunck},\ and\ \citenamefont
  {Stetcu}}]{Bulgac:2019b}%
  \BibitemOpen
  \bibfield  {author} {\bibinfo {author} {\bibfnamefont {A.}~\bibnamefont
  {Bulgac}}, \bibinfo {author} {\bibfnamefont {S.}~\bibnamefont {Jin}},
  \bibinfo {author} {\bibfnamefont {K.~J.}\ \bibnamefont {Roche}}, \bibinfo
  {author} {\bibfnamefont {N.}~\bibnamefont {Schunck}}, \ and\ \bibinfo
  {author} {\bibfnamefont {I.}~\bibnamefont {Stetcu}},\ }\bibfield  {title}
  {\enquote {\bibinfo {title} {Fission dynamics of $^{240}\mathrm{Pu}$ from
  saddle to scission and beyond},}\ }\href {\doibase
  10.1103/PhysRevC.100.034615} {\bibfield  {journal} {\bibinfo  {journal}
  {Phys. Rev. C}\ }\textbf {\bibinfo {volume} {100}},\ \bibinfo {pages}
  {034615} (\bibinfo {year} {2019}{\natexlab{b}})}\BibitemShut {NoStop}%
\bibitem [{\citenamefont {Sekizawa}(2017)}]{Sekizawa:2017a}%
  \BibitemOpen
  \bibfield  {author} {\bibinfo {author} {\bibfnamefont {K.}~\bibnamefont
  {Sekizawa}},\ }\bibfield  {title} {\enquote {\bibinfo {title} {Microscopic
  description of production cross sections including deexcitation effects},}\
  }\href {\doibase 10.1103/PhysRevC.96.014615} {\bibfield  {journal} {\bibinfo
  {journal} {Phys. Rev. C}\ }\textbf {\bibinfo {volume} {96}},\ \bibinfo
  {pages} {014615} (\bibinfo {year} {2017})}\BibitemShut {NoStop}%
\bibitem [{\citenamefont {Bulgac}(2019{\natexlab{b}})}]{Bulgac:2019d}%
  \BibitemOpen
  \bibfield  {author} {\bibinfo {author} {\bibfnamefont {A.}~\bibnamefont
  {Bulgac}},\ }\bibfield  {title} {\enquote {\bibinfo {title} {Projection of
  good quantum numbers for reaction fragments},}\ }\href {\doibase
  10.1103/PhysRevC.100.034612} {\bibfield  {journal} {\bibinfo  {journal}
  {Phys. Rev. C}\ }\textbf {\bibinfo {volume} {100}},\ \bibinfo {pages}
  {034612} (\bibinfo {year} {2019}{\natexlab{b}})}\BibitemShut {NoStop}%
\bibitem [{\citenamefont {Ring}\ and\ \citenamefont
  {Schuck}(2004)}]{Ring:2004}%
  \BibitemOpen
  \bibfield  {author} {\bibinfo {author} {\bibfnamefont {P.}~\bibnamefont
  {Ring}}\ and\ \bibinfo {author} {\bibfnamefont {P.}~\bibnamefont {Schuck}},\
  }\href@noop {} {\emph {\bibinfo {title} {{The Nuclear Many-Body Problem}}}},\
  \bibinfo {edition} {1st}\ ed.,\ \bibinfo {series} {Theoretical and
  Mathematical Physics Series}\ No.~\bibinfo {number} {17}\ (\bibinfo
  {publisher} {Springer-Verlag},\ \bibinfo {address} {Berlin Heidelberg New
  York},\ \bibinfo {year} {2004})\BibitemShut {NoStop}%
\bibitem [{\citenamefont {Bertsch}\ \emph {et~al.}(2019)\citenamefont
  {Bertsch}, \citenamefont {Kawano},\ and\ \citenamefont
  {Robledo}}]{Bertsch:2019}%
  \BibitemOpen
  \bibfield  {author} {\bibinfo {author} {\bibfnamefont {G.~F.}\ \bibnamefont
  {Bertsch}}, \bibinfo {author} {\bibfnamefont {T.}~\bibnamefont {Kawano}}, \
  and\ \bibinfo {author} {\bibfnamefont {L.~M.}\ \bibnamefont {Robledo}},\
  }\bibfield  {title} {\enquote {\bibinfo {title} {Angular momentum of fission
  fragments},}\ }\href {\doibase 10.1103/PhysRevC.99.034603} {\bibfield
  {journal} {\bibinfo  {journal} {Phys. Rev. C}\ }\textbf {\bibinfo {volume}
  {99}},\ \bibinfo {pages} {034603} (\bibinfo {year} {2019})}\BibitemShut
  {NoStop}%
\bibitem [{\citenamefont {Scamps}\ and\ \citenamefont
  {Simenel}(2018)}]{Scamps:2018}%
  \BibitemOpen
  \bibfield  {author} {\bibinfo {author} {\bibfnamefont {G.}~\bibnamefont
  {Scamps}}\ and\ \bibinfo {author} {\bibfnamefont {C.}~\bibnamefont
  {Simenel}},\ }\bibfield  {title} {\enquote {\bibinfo {title} {Impact of
  pear-shaped fission fragments on mass-asymmetric fission in actinides},}\
  }\href {\doibase 10.1038/s41586-018-0780-0} {\bibfield  {journal} {\bibinfo
  {journal} {Nature}\ }\textbf {\bibinfo {volume} {564}},\ \bibinfo {pages}
  {382} (\bibinfo {year} {2018})}\BibitemShut {NoStop}%
\bibitem [{\citenamefont {Strutinsky}(1967)}]{Strutinsky:1967}%
  \BibitemOpen
  \bibfield  {author} {\bibinfo {author} {\bibfnamefont {V.M.}\ \bibnamefont
  {Strutinsky}},\ }\bibfield  {title} {\enquote {\bibinfo {title} {Shell
  effects in nuclear masses and deformation energies},}\ }\href {\doibase
  https://doi.org/10.1016/0375-9474(67)90510-6} {\bibfield  {journal} {\bibinfo
   {journal} {Nucl. Phys. A}\ }\textbf {\bibinfo {volume} {95}},\ \bibinfo
  {pages} {420} (\bibinfo {year} {1967})}\BibitemShut {NoStop}%
\bibitem [{\citenamefont {Brack}\ \emph {et~al.}(1972)\citenamefont {Brack},
  \citenamefont {Damgaard}, \citenamefont {Jensen}, \citenamefont {Pauli},
  \citenamefont {Strutinsky},\ and\ \citenamefont {Wong}}]{BRACK:1972}%
  \BibitemOpen
  \bibfield  {author} {\bibinfo {author} {\bibfnamefont {M.}~\bibnamefont
  {Brack}}, \bibinfo {author} {\bibfnamefont {J.}~\bibnamefont {Damgaard}},
  \bibinfo {author} {\bibfnamefont {A.~S.}\ \bibnamefont {Jensen}}, \bibinfo
  {author} {\bibfnamefont {H.~C.}\ \bibnamefont {Pauli}}, \bibinfo {author}
  {\bibfnamefont {V.~M.}\ \bibnamefont {Strutinsky}}, \ and\ \bibinfo {author}
  {\bibfnamefont {C.~Y.}\ \bibnamefont {Wong}},\ }\bibfield  {title} {\enquote
  {\bibinfo {title} {{Funny Hills: The Shell-Correction Approach to Nuclear
  Shell Effects and Its Applications to the Fission Process}},}\ }\href
  {\doibase 10.1103/RevModPhys.44.320} {\bibfield  {journal} {\bibinfo
  {journal} {Rev. Mod. Phys.}\ }\textbf {\bibinfo {volume} {44}},\ \bibinfo
  {pages} {320} (\bibinfo {year} {1972})}\BibitemShut {NoStop}%
\bibitem [{\citenamefont {Bohr}\ and\ \citenamefont
  {Mottelson}(1969)}]{Bohr:1969}%
  \BibitemOpen
  \bibfield  {author} {\bibinfo {author} {\bibfnamefont {A.}~\bibnamefont
  {Bohr}}\ and\ \bibinfo {author} {\bibfnamefont {B.~R.}\ \bibnamefont
  {Mottelson}},\ }\href@noop {} {\emph {\bibinfo {title} {Nuclear Structure}}}\
  (\bibinfo  {publisher} {Benjamin Inc.},\ \bibinfo {address} {New York},\
  \bibinfo {year} {1969})\BibitemShut {NoStop}%
\bibitem [{\citenamefont {Marevic}\ \emph {et~al.}(2020)\citenamefont
  {Marevic}, \citenamefont {Schunck},\ and\ \citenamefont {anmd
  R.~Vogt}}]{Schunck:2020x}%
  \BibitemOpen
  \bibfield  {author} {\bibinfo {author} {\bibfnamefont {P.}~\bibnamefont
  {Marevic}}, \bibinfo {author} {\bibfnamefont {N.}~\bibnamefont {Schunck}}, \
  and\ \bibinfo {author} {\bibfnamefont {J.~Randrup}\ \bibnamefont {anmd
  R.~Vogt}},\ }\href@noop {} {\enquote {\bibinfo {title} {{Angular Momentum of
  Fission Fragments from Microscopic Theory}},}\ } (\bibinfo {year} {2020}),\
  \Eprint {http://arxiv.org/abs/2101.03406} {arXiv:2101.03406} \BibitemShut
  {NoStop}%
\end{thebibliography}%

\end{document}